\documentclass[showpacs,twocolumn,preprintnumbers,amsmath,amssymb,pre,floatfix]{revtex4}

\usepackage[english]{babel}
\usepackage[latin2]{inputenc}
\usepackage[T1]{fontenc}

\usepackage{ae,aecompl}
\usepackage{enumerate}
\usepackage{epsfig}
\usepackage{graphics}
\usepackage{graphicx}
\usepackage{amsmath}
\usepackage{amssymb}
\usepackage{pifont}
\usepackage[sort&compress]{natbib}
\usepackage{wasysym}

\def\ccc#1;#2{\left\langle #1 \left\vert #2 \right.\right\rangle}
\def\ev #1{\left\langle #1 \right\rangle}

\begin{document}

\preprint{}
\title{Scaling theory of temporal correlations and size dependent fluctuations in the traded value of stocks}
\author{Zolt\'an Eisler}
\email{eisler@maxwell.phy.bme.hu}
\author{J\'anos Kert\'esz}
\altaffiliation[Also at ]{Laboratory of Computational Engineering, Helsinki
University of Technology, Espoo, Finland} \affiliation{Department of Theoretical
Physics, Budapest University of Technology and Economics, Budapest, Hungary}
\date{\today}


\begin{abstract}
Records of the traded value $f_i$ of stocks display fluctuation scaling, a
proportionality between the standard deviation $\sigma_i$ and the average
$\left\langle f_i\right\rangle$: $\sigma_i \propto
\left\langle f_i\right\rangle^\alpha$, with a strong time scale dependence
$\alpha(\Delta t)$. The non-trivial (i.e., neither $0.5$ nor $1$) value of
$\alpha$ may have different origins and provides information about the
microscopic dynamics. We present a set of new stylized facts,
and then show their connection to such behavior. The functional form
$\alpha(\Delta t)$ originates from two aspects of the dynamics: Stocks of
larger companies both tend to be traded in larger packages, and also display
stronger correlations of traded value. The results are integrated into a
general framework that can be applied to a wide range of complex systems.
\end{abstract}

\maketitle

\section{Introduction}
Research concerning the forces that govern stock markets is
largely helped by the abundant availability of data on trading
activity \cite{bouchaud.book, stanley.book}. Recently an increasing
number of complex systems have been studied in a similar,
data-oriented way \cite{disordered, barabasi.rmp}. Examples include records of
information flow through routers on the Internet, or of web page
visitations \cite{barabasi.fluct, barabasi.separating}. The means to
extract information from such multichannel data are generally limited
to the system-wide distributions of various quantities and
cross-correlation measurements. Although these approaches have been
very fruitful in various areas, they often fail to provide information
on the mechanisms that govern the observed internal processes. On the
grounds of a recently discovered set of empirical facts regarding
stock market trading activity \cite{eisler.non-universality,
eisler.sizematters, ivanov.itt, ivanov.unpublished}, we test a new
tool that addresses these questions. It is based
on an empirical scaling law that appears to hold for most systems. It
connects the fluctuations $\sigma_i$ and the average $\ev{f_i}$ of the
activity of constituents by a \emph{fluctuation scaling} law:
$$\sigma_i \propto \ev{f_i}^\alpha.$$

In several studies, the value of $\alpha$ is used as a proxy
of the dominant factors of internal dynamics \cite{barabasi.fluct, barabasi.separating,
eisler.non-universality}. $\alpha = 0.5$ is said to characterize equilibrium systems,
while $\alpha = 1$ is considered  a consequence of strong external driving.
While the general idea works well in a number of settings, we find that $\alpha$
shows a much richer behavior than previously anticipated. When calculated for systems with strong temporal correlations,
$\alpha$ becomes time scale dependent. Moreover, even in the lack of correlations, $\alpha > 0.5$
is possible. The aim of this paper is twofold: We present a detailed theory for the emergence of fluctuation scaling, which explains such anomalies, and we also apply this theory to explain phenomena observed in finance.

Section \ref{sec:notations} introduces notations and our set of stock market
data. Section \ref{sec:stylized} presents some new stylized
facts regarding stock market trading activity, and then Section
\ref{sec:scaling} describes the concept of fluctuation scaling that connects
all those observations, including the previously identified two universality classes:
$\alpha = 0.5$ and $\alpha = 1$. Then, we deal with a mechanism that explains
how stock markets can display a non-universal value of $\alpha \approx 0.68$.
Finally, we describe how dynamical correlations are reflected in the time scale
dependence of the exponent $\alpha$.

\section{Notations and data}
\label{sec:notations}
For our analysis of financial data, it is necessary to give a few definitions.
For a time window size $\Delta t$, one can write the total traded value of the
$i$th stock at time $t$ in the form
\begin{equation}
f_i^{\Delta t}(t) = \sum_{n, t_i(n)\in [t, t+\Delta t]} V_i(n),
\label{eq:flow}
\end{equation}
where $t_i(n)$ is the time of the $n$-th transaction of stock $i$. The
so called tick-by-tick data is denoted by $V_i(n)$, this is the value traded in
transaction $n$. It can be calculated as the product of the price $p$ and the
traded volume $\tilde V$:
\begin{equation}
V_i(n) = p_i(n) \tilde V_i(n).
\label{eq:v}
\end{equation}
The price serves as a weighting factor to make the comparison of
different stocks possible, while this definition also eliminates the
effect of stock splits.

As the source of empirical data, we used the TAQ database
\cite{taq2000-2002}, that records all transactions of the New York
Stock Exchange (NYSE) and NASDAQ for the years $2000-2002$. Our sample
was restricted to those $2647$ stocks for NYSE and $4039$ for NASDAQ, that were
continuously traded in the period. We divided the data by the well-known $U$-shaped daily
pattern of traded volumes, similarly to Ref. \cite{eisler.non-universality}.

Finally, note that we use $10$-base logarithms throughout the paper to
ease the understanding of financial data.

\section{Stylized facts of trading activity: Summary and new results}
\label{sec:stylized}
This section presents a few recent advances 
in understanding the empirical properties of trading activity. Their
focus is on the fundamental role of company size. 
This is usually measured by the capitalization, but
that is closely related to the trading frequency, which in turn
influences a very broad range of statistical properties observed in
data.

\subsection{Size-dependent correlations}
The presence of long-range autocorrelations in various measures of trading is a
well-known fact \cite{eisler.sizematters, ivanov.itt, ivanov.unpublished}. For
example, stock market volatility
\cite{bouchaud.book, stanley.book, cont.stylized} and trading volumes
\cite{gopi.volume,
eisler.sizematters} show strong persistence. Correlations can be
characterized by the Hurst exponent $H(i)$ \cite{vicsek.book,
dfa}. For stock $i$, this is defined \footnote{Despite earlier arguments
\cite{gopi.volume}, $\sigma_i(\Delta t)$ is not divergent \cite{queiros.volume,
eisler.sizematters} and so $H(i)$ can indeed be introduced.} as
\begin{equation}
\sigma_i(\Delta t) = \ev{\left ( f_i^{\Delta t}(t) - \ev{f_i^{\Delta t}(t)}
\right )^2} \propto \Delta t^{H(i)}, \label{eq:hurst}
\end{equation}
where $\ev{\cdot}$ denotes time averaging. There is a range of methods
\cite{dfa, dfa.intro, muzy.wtmm} to estimate the Hurst exponent, and
the understanding of the results is well established
\cite{vicsek.book}. The signal is said to be correlated (persistent)
when $H>0.5$, uncorrelated when $H=0.5$, and anticorrelated
(antipersistent) for $H<0.5$.

\begin{figure}[tbp]
\centerline{\includegraphics[height=190pt]{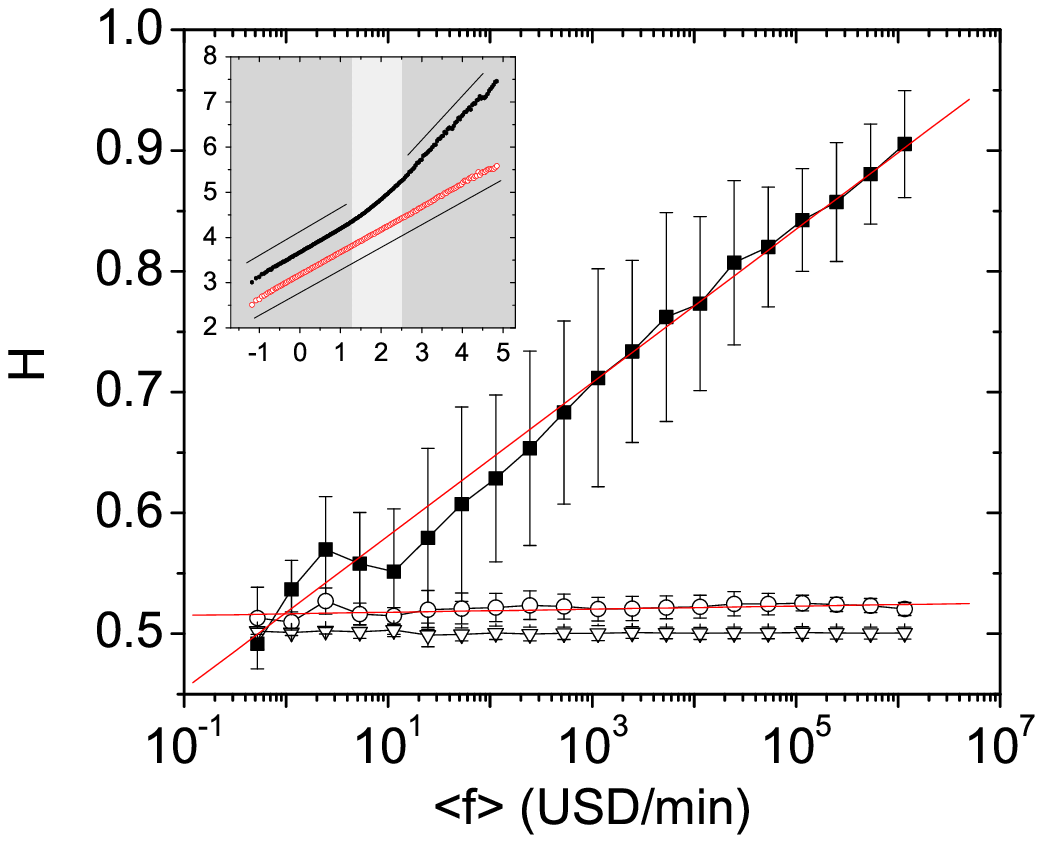}}
\caption{(color online) Behavior of the Hurst exponents $H(i)$ for NYSE stocks in the
period $2000-2002$. For short time windows ($\Circle$), all signals
are nearly uncorrelated, $H(i)\approx 0.51 - 0.52$. The fitted slope
is $\gamma_{t}(\Delta t < \mathrm{20\mathrm{\space min}})=0.001\pm 0.002$. For
larger time windows ($\blacksquare$), the strength of correlations
depends logarithmically on the mean trading activity of the stock,
$\gamma_{t}(\Delta t > \mathrm{300\mathrm{ min}})=0.06\pm 0.01$. Shuffled
data ($\bigtriangledown$) display no correlations, thus $H_\mathrm{shuff}(i) = 0.5$, which also
implies $\gamma_{t} = 0$. The inset shows the
$\log \sigma$-$\log \Delta t$ scaling plot for General Electric (GE). The
darker shaded
intervals have well-defined Hurst exponents, the crossover is indicated with a
lighter background.
The slopes corresponding to Hurst exponents are $0.53$ and $0.93$; the
slope for shuffled data is $0.51$. Shuffled points were shifted
vertically for better visibility.}
\label{fig:hurstNYSE}
\end{figure}

\begin{figure}[tbp]
\centerline{\includegraphics[height=190pt]{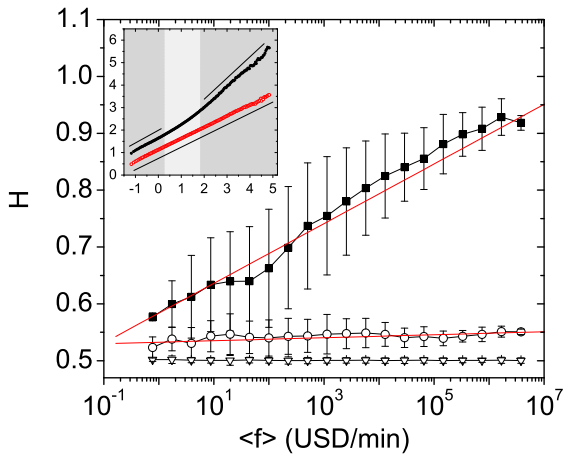}}
\caption{(color online) Behavior of the Hurst exponents $H(i)$ for NASDAQ stocks in
the period $2000-2002$. For short time windows ($\Circle$), all
signals are nearly uncorrelated, $H\approx 0.52 - 0.53$. The fitted
slope is $\gamma_{t}(\Delta t < \mathrm{2\mathrm{ min}})=0.003\pm 0.002$. For
larger time windows ($\blacksquare$), the strength of correlations
depends logarithmically on the mean trading activity of the stock,
$\gamma_{t}(\Delta t > \mathrm{60\mathrm{ min}})=0.05\pm 0.01$.  Shuffled
data ($\bigtriangledown$) display no correlations, thus $H_\mathrm{shuff}(i) = 0.5$, which
also implies $\gamma_{t} = 0$. The inset shows the
$\log \sigma$-$\log \Delta t$ scaling plot for Dell (DELL). The darker shaded
intervals have well-defined Hurst exponents, the crossover is indicated with a
lighter background. The slopes corresponding to Hurst exponents are $0.54$ and
$0.90$; the slope for shuffled data is $0.50$. Shuffled points were
shifted vertically for better visibility.}
\label{fig:hurstNASDAQ}
\end{figure}

It is intriguing, that stock market activity has a much richer
behavior, than simply all stocks having Hurst exponents statistically
distributed around an average value, as assumed in 
Ref. \cite{gopi.volume}. Instead, there is a crossover
\cite{eisler.sizematters, ivanov.itt, ivanov.unpublished} between two
types of behavior around the time scale of $1$ day. We located this
threshold by a technique that will be discussed in Section
\ref{sec:correl}. An essentially uncorrelated regime was found when
$\Delta t < 20$ min for NYSE and $\Delta t < 2$ min for NASDAQ, while
the time series of larger companies become strongly correlated when
$\Delta t > 300$ min for NYSE and $\Delta t > 60$ min for NASDAQ. As a
reference, we also calculated the Hurst exponents $H_{\mathrm{shuff}}(i)$ of
the shuffled time series. Results are given in
Figs. \ref{fig:hurstNYSE} and \ref{fig:hurstNASDAQ} \footnote{We also investigated the effect of randomly shuffling the Fourier-phases of the data, which destroys the possible non-linearities in the time series. One finds, that the crossover behavior persists after such a transformation.}.

One can see, that for shorter time windows, correlations are absent in both
markets, $H(i)\approx0.51-0.53$. For windows longer than a trading day, however,
while small $\ev{f}$ stocks
again display only very weak correlations, larger ones show up to $H\approx
0.9$. Furthermore, there is a distinct logarithmic trend in the data:
\begin{equation}
H(i) = H^* + \gamma_{t}\log\ev{f_i},
\end{equation}
with $\gamma_{t}(\Delta t > 300\mathrm{ min}) = 0.06\pm0.01$ for NYSE and
$\gamma_{t}(\Delta t > 60\mathrm{ min}) = 0.05\pm0.01$
for NASDAQ. Shorter time scales correspond to the special case $\gamma_{t} = 0$,
there is no systematic trend in $H$. Shuffled data, as expected,
show $H_\mathrm{shuff}(i)\approx 0.5$ at all time scales and without significant
dependence on $\ev{f_i}$.

It is to be emphasized, that the crossover is not simply between uncorrelated
and correlated regimes. It is instead between homogeneous (all stocks
show $H(i)\approx H_1$, $\gamma_{t} = 0$) and inhomogeneous ($\gamma_{t} > 0$)
behavior. One finds $H_1 \approx 0.5$, but very small
$\ev{f}$ stocks do not depart much from this value even for large time
windows. This is a clear relation to company size, as $\ev{f}$
is a monotonically growing function of company capitalization
\cite{eisler.sizematters}. Dependence of the effect on $\ev{f}$ is in fact a
dependence on company size.

\subsection{Fluctuation scaling of $f$}

\begin{figure}[ptb]
\centerline{\includegraphics[height=160pt]{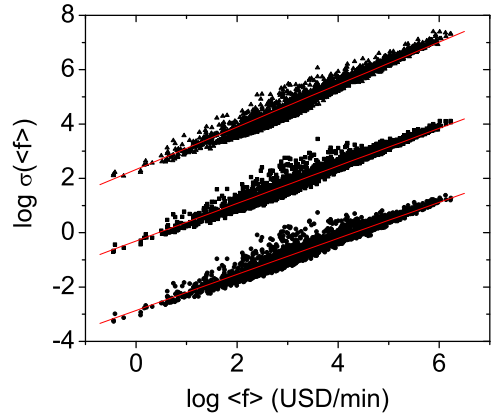}}
\caption{(color online) Examples of $\sigma(\ev{f})$ scaling plots for NYSE, years
$2000-2002$. The window sizes from bottom to top: $\Delta t = 10$ sec,
$0.5$ day, $2$ weeks. The slopes are $\alpha = 0.68, 0.71, 0.80$,
respectively. Points were shifted vertically for better visibility.}
\label{fig:alphaexampleNYSE}
\end{figure}

This paper will mainly focus on a special property of the time series
$f_i^{\Delta t}(t)$: \emph{fluctuation scaling}
\cite{barabasi.fluct, barabasi.separating, eisler.non-universality}.
This connects the standard deviation $\sigma_i$ and the average $\ev{f_i}$
of the trading activity for all our $i=1\dots N$ stocks:
\begin{equation}
\sigma_i(\Delta t) \propto \ev{f_i}^{\alpha (\Delta t)}.
\label{eq:alpha}
\end{equation}
Due to the long time period ($3$ years), the data are highly
instationary. Thus, unlike a previous study \cite{eisler.non-universality},
here we applied the DFA procedure \cite{dfa, dfa.intro}
to estimate $\sigma_i(\Delta t)$. We determined the values of
$\alpha$ for traded value fluctuations by fits to \eqref{eq:alpha},
examples are shown in Fig. \ref{fig:alphaexampleNYSE}.

\begin{figure*}[ptb]
\centerline{\includegraphics[height=175pt]{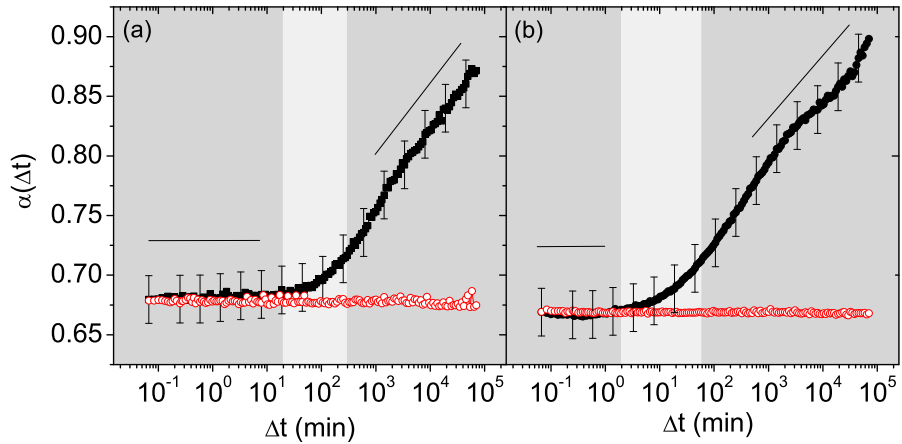}}
\caption{(color online) The dependence of the scaling exponent $\alpha$ on the window
size $\Delta t$. The darker shaded intervals have well-defined Hurst exponents
and values of $\gamma_{t}$, the crossover is indicated with a lighter background.
{\bf (a)} NYSE: without shuffling ($\blacksquare$)
the slopes of the linear regimes are $\gamma_{f}(\Delta t<20$
min$)=0.00\pm 0.01$ and $\gamma_{f}(\Delta t>300$ min$)=0.06\pm 0.01$. For
shuffled data ($\Circle$) the exponent is independent of window size,
$\alpha (\Delta t)=0.68\pm0.02$. {\bf (b)} NASDAQ: without shuffling
($\blacksquare$) the slopes of the linear regimes are $\gamma_{f}(\Delta
t<2$ min$)=0.00\pm0.01$ and $\gamma_{f}(\Delta t > 60$
min$)=0.06\pm0.01$. For shuffled data ($\Circle$) the exponent is
independent of window size, $\alpha (\Delta t)=0.67\pm0.02$.
\emph{Note}: There is a deviation from linearity around $\Delta t \approx 1$
trading week. It is larger for NASDAQ, but it is still between the error bars.
A possible cause is the weekly periodic pattern of trading which was not removed manually.}
\label{fig:alpha}
\end{figure*}

The exponent $\alpha$ strongly depends on the size $\Delta t$ of the time
windows. Recently, Refs.
\cite{zawa.pricechanges1, zawa.pricechanges2, ivanov.unpublished} pointed out
that the trading activity of NYSE and NASDAQ display very different
temporal correlations, possibly due to their different
trading mechanisms. Still, the scaling \eqref{eq:alpha} does
hold regardless of market and $\Delta t$. Furthermore, the functions $\alpha(\Delta t)$ agree
qualitatively. The exponents are shown for NYSE and NASDAQ in
Figs. \ref{fig:alpha}(a) and \ref{fig:alpha}(b), respectively. One can
see, that $\alpha$ is a non-decreasing function of $\Delta t$, and in
large regimes it is, to a good approximation, either constant or
logarithmic.

\subsection{Fluctuation scaling of $N$ and $V$}

One can carry out a similar analysis of other quantities, here we limit ourself
to two of those. The first one, the number of trades of stock $i$ in size
$\Delta t$ time windows, will be denoted by $N_i^{\Delta t}(t)$, its variance
by $\sigma_N^2(i, \Delta t)$. The second one was introduced before, $V_i(n)$
is the value exchanged in the $n$'th trade of stock $i$. The corresponding
variance will be $\sigma^2_{Vi}$.

\begin{figure}[tbp]
\includegraphics[height=160pt]{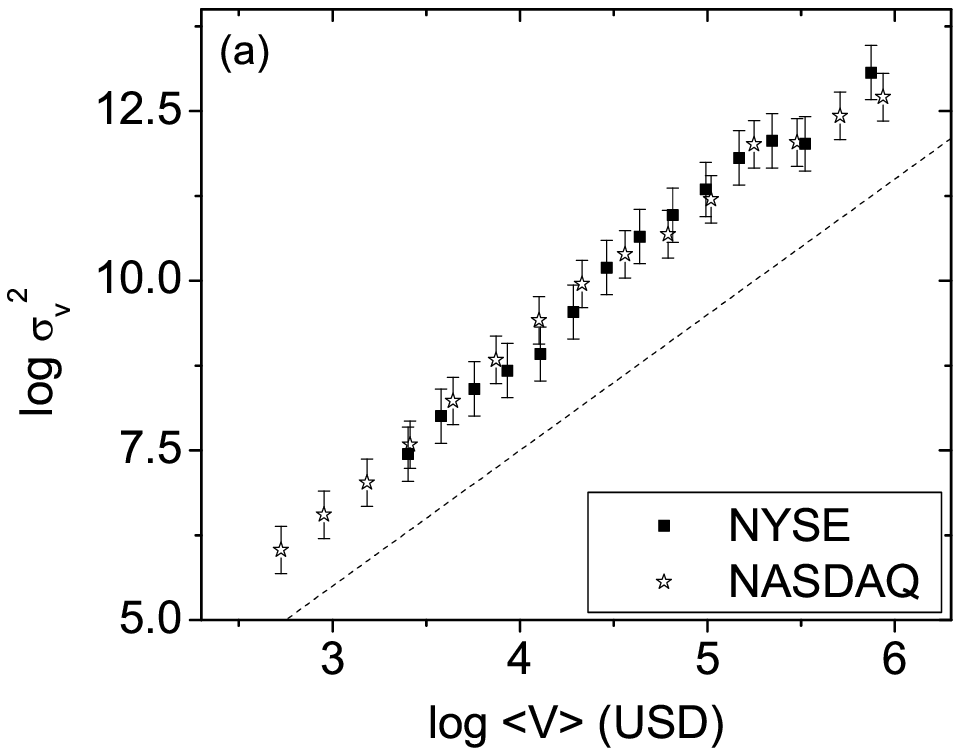}
\includegraphics[height=160pt]{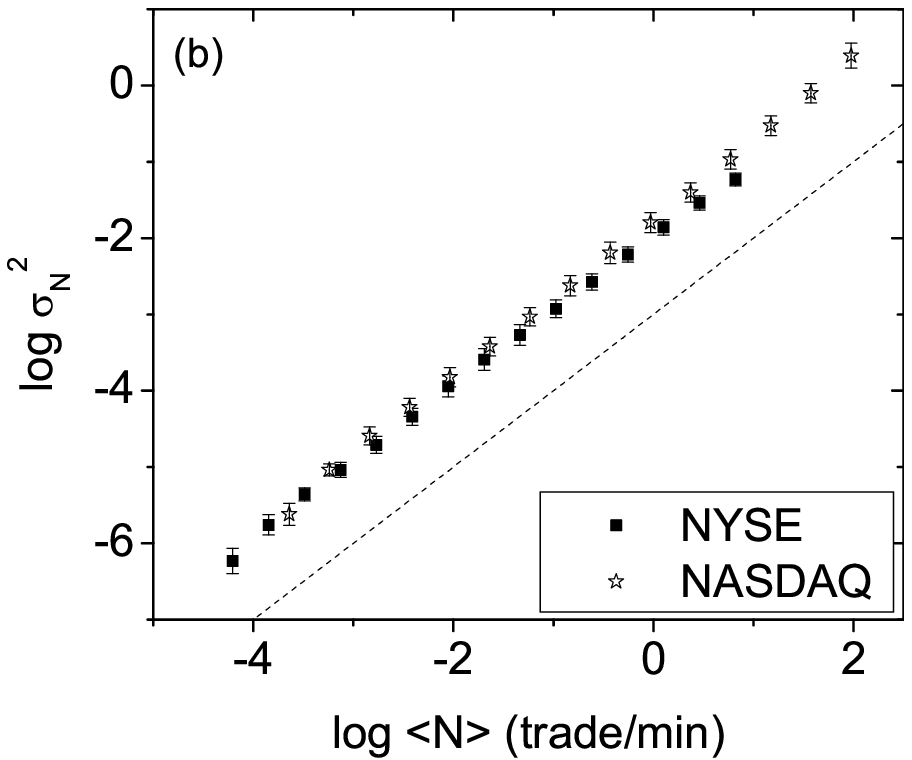}
\caption{{\bf (a)} Plot verifying the validity of \eqref{eq:s2V} for stock
market data, typical error bars are given. The straight line would correspond
to $\sigma^2_{Vi} \propto \ev{V_i}^2$. {\bf (b)} Plot verifying the validity of
\eqref{eq:s2N} for stock market data, typical error bars are given. The straight
line would correspond to $\sigma^2_{Ni} \propto \ev{N_i}$.
The size of the time windows is $\Delta t = 1$ sec.} 
\label{fig:conditions}
\end{figure}

Dimensional analysis predicts
\begin{equation}
\sigma^2_{Vi} \propto \ev{V_i}^2, 
\label{eq:s2V}
\end{equation}
which is remarkably close to the observed behavior, shown in Fig.
\ref{fig:conditions}(a). Also, when the size of the time windows is chosen
sufficiently small ($\Delta t \ll 1$ min), the probability that two
trades of the same stock happen in the same period is negligible. In this limit,
correlations between consecutive trades cannot contribute to $\sigma^2_N$, the
central limit theorem becomes applicable, and simply
\begin{equation}
\sigma^2_{Ni} \propto \ev{N_i},
\label{eq:s2N}
\end{equation}
which again agrees very well with empirical data shown for $\Delta t = 1$ sec
in Fig. \ref{fig:conditions}(b).

\subsection{Dependence of typical trade size on trading frequency}

\begin{figure}[tbp]
\includegraphics[height=160pt]{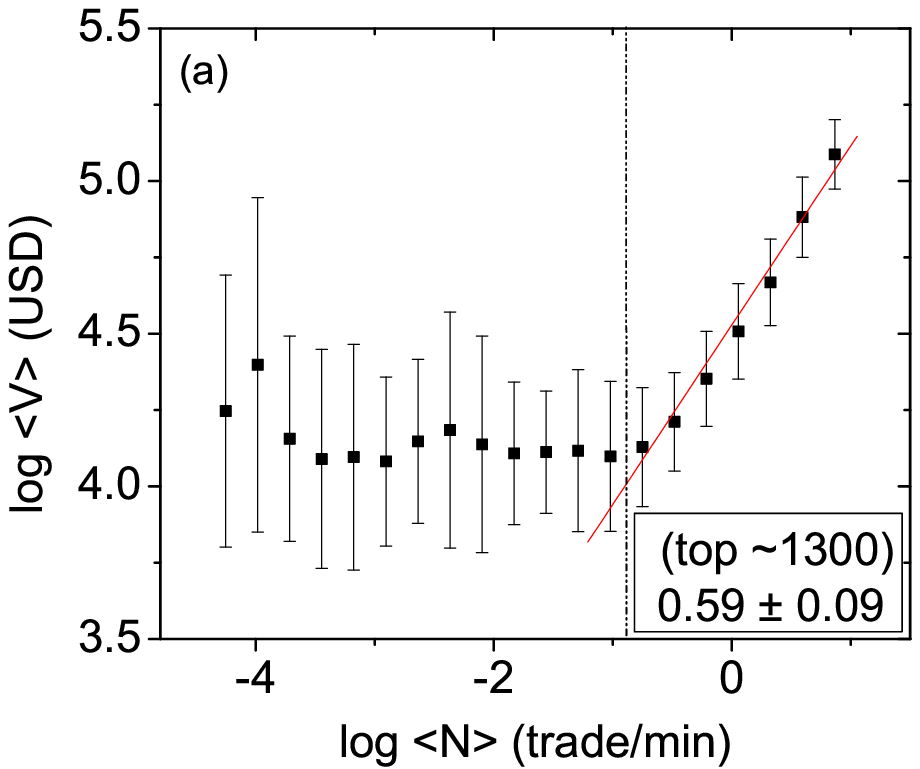}
\includegraphics[height=160pt]{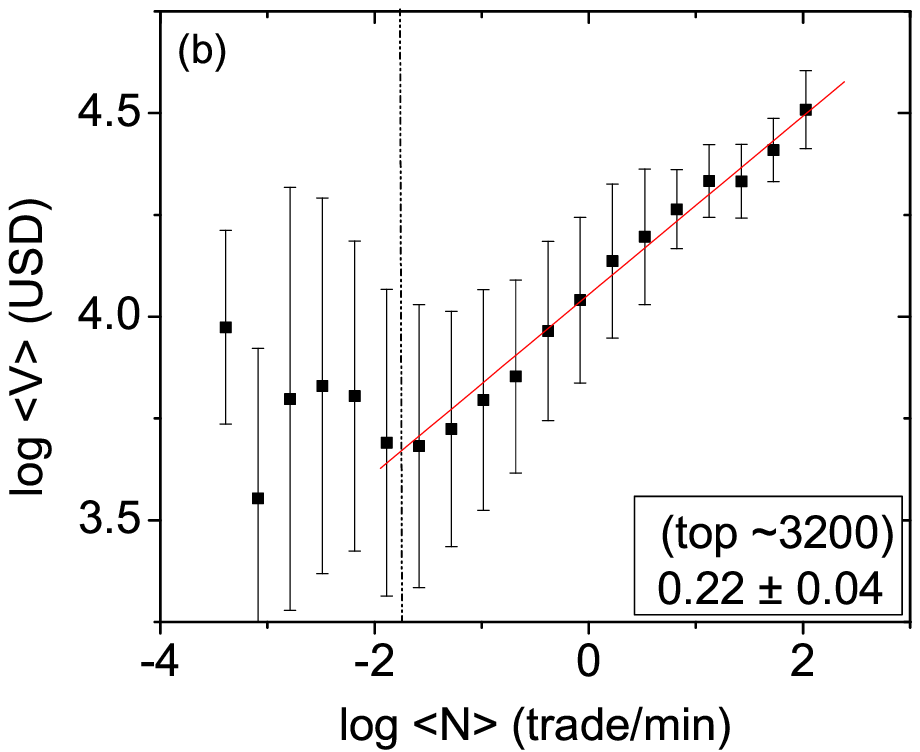}
\caption{(color online) The dependence of the mean value per trade $\ev{V_i}$ on the
average rate of trades $\ev{N_i}$. Calculations were done for the
period $2000-2002$, {\bf (a)} shows NYSE and {\bf (b)} shows
NASDAQ. Points were binned and their logarithm was averaged for better
visibility, error bars show the standard deviations in the bins. For
the smallest stocks there is no clear trend at either
exchange. However, larger stocks at NYSE and all except the minor ones
at NASDAQ, show scaling between the two quantities, equivalent to that
given in \eqref{eq:vvsn}. The slopes are $\beta_\mathrm{NYSE} = 0.59 \pm
0.04$ and $\beta_\mathrm{NASDAQ} = 0.22\pm 0.04$.}
\label{fig:mNvsmpV}
\end{figure}

The final observation to be discussed here is that for
a large group of stocks, the average rate of trades $\ev{N}$ and their
mean value $\ev{V}$ are connected by a power law:
\begin{equation}	
\ev{V_i} \propto \ev{N_i}^\beta .
\label{eq:vvsn}
\end{equation}
Such relationships are shown in Figs. \ref{fig:mNvsmpV}(a) and
\ref{fig:mNvsmpV}(b) for NYSE and
NASDAQ, respectively. The measured exponents are $\beta_\mathrm{NYSE} = 0.59
\pm 0.09$ and $\beta_\mathrm{NASDAQ} = 0.22 \pm 0.04$, although they are
restricted to large enough stocks. The estimate based on Ref. \cite{zumbach}
for the stocks in London's FTSE-100, is $\beta \approx 1$.

The values of $\beta_\mathrm{NYSE}$ and $\beta_\mathrm{NASDAQ}$, and especially the marked difference between them appears to be very robust for various time periods. One major contribution to this is probably the difference in trading mechanisms between the two markets \cite{uponrequest, ivanov.unpublished}.

One very crude interpretation of the effect in general is the following. Smaller stocks are exchanged
rarely, but transaction costs must limit from below the value that is still
profitable to be exchanged at once. This minimal unit is around the order of
$10^4$ USD for both markets. Once the speed of trading and liquidity grow,
it becomes possible to exchange larger packages. Trades start to
"stick together", their average value starts to grow. Although this tendency
reduces transaction costs, the price impact
\cite{gabaix.powerlaw, plerou.powerlaw, farmer.powerlaw, farmer.whatreally} of
the trade also increases, which in practice often limits package sizes from
above. These two mechanisms may have a role in the formation of \eqref{eq:vvsn}.
Also, as they vary strongly from market to market, such very different values of
$\beta$ might be justified.

\section{Scaling theory}
\label{sec:scaling}
In this section, we present a framework that unifies the --
seemingly unrelated -- observations of Sec. \ref{sec:stylized}. This is centered around the above
introduced fluctuation scaling \eqref{eq:alpha}:
$$\sigma_i(\Delta t) \propto \ev{f_i}^{\alpha (\Delta t )}.$$
This phenomenon is not at all specific to stock market data, in fact it has
been observed for activity in a wide range of complex systems. Possible choices
for $f$ include data traffic through Internet routers, daily web page hits,
highway traffic \cite{barabasi.fluct, barabasi.separating} and node visitations
of random walkers on complex networks \cite{barabasi.fluct, eisler.internal}.
In this sense, the stock market is seen as a complex system, where the constituents are
stocks and their activity at any time is given by the traded value per unit
time.

\subsection{Universal values of $\alpha$}
\label{sec:universal}
First, notice that \eqref{eq:hurst} and \eqref{eq:alpha} are formal
analogues. They connect the same standard deviation with the two
complementary factors: the $\Delta t$ size of the time window and the average
(trading) activity $\ev{f_i}$. There is evidence, that while $H(i)$ describes
the correlation properties of the individual elements activity, the function
$\alpha (\Delta t)$ carries information about the collective dynamical
properties of the whole system. Based on this knowledge, a
classification scheme was outlined in Refs.
\cite{barabasi.fluct, eisler.internal, eisler.nikkei}.
All those studies assume, that the activities of all nodes are uncorrelated,
i.e., $H(i)=0.5$ \footnote{Note that in general, instead of uncorrelated
dynamics, it is enough if the activity of every node displays the same
Hurst exponent, $H(i)=H$. This is the direct consequence of arguments
in Section \ref{sec:scaling}.}. In this case, there are two known universality
classes with respect to the value of $\alpha$.

In certain systems, the activity of the constituents comes from nearly
equivalent, independent events. The difference between nodes with
smaller and greater mean activity comes basically from the different
mean \emph{number} of events. Then, the \emph{central limit theorem}
can be applied to these events and this yields $\alpha = 0.5$
automatically. Examples include simple surface growth models and the
data traffic of Internet routers \cite{barabasi.fluct}.

Other systems dynamics is under a \emph{dominant external driving
force}: Activity fluctuations are mainly caused by the variations of
this external force, and this leads to proportionality between the strength
and the standard deviations at the nodes: $\alpha = 1$, regardless of
the internal structure or the laws governing the time evolution. This is
observed for the statistics of web page visitations and highway traffic
\cite{barabasi.fluct}.

In temporally uncorrelated systems, two processes are known to give
rise to intermediate measured values $0.5 < \alpha < 1$: Some finite
systems display a crossover between $\alpha = 0.5$ and $\alpha = 1$ at
a certain node strength $\ev{f}$, due to the competition of external
driving and internal dynamics \cite{barabasi.fluct, barabasi.separating}.
There is an \emph{effective} value of $\alpha$, but in fact, scaling breaks
down. Another possible scenario is discussed in the following. 

\subsection{Non-universal values of $\alpha$}
\label{sec:internal}
The activities $f_i(t)$ originate from individual events that take 
place at the nodes. Every event $n$ at node $i$ is characterized by its time
$t_i(n)$ and its size $V_i(n)$ which is now allowed to vary. For a given size of time windows $\Delta t$, the observed time series is given by
$$f_i^{\Delta t}(t) = \sum_{n, t_i(n)\in [t, t+\Delta t]} V_i(n),$$
a formula equivalent to \eqref{eq:flow}. In the stock market, the value
exchanged in a trade is a plausible choice of $V$.
 
If the random process that gives the size of an event is independent of the
one that determines when the event occurs, one can find a simple formula
\cite{eisler.internal} that shows how fluctuations of $f$ are composed:
\begin{equation}
\sigma^2_i = \sigma^2_{Vi} \ev{N_i} + \sigma^2_{Ni} \ev{V_i}^2,
\end{equation}
where $\ev{V_i}$ and $\sigma^2_{Vi}$ are the mean and the standard
deviation of the event size distribution. $\ev{N_i}$ and
$\sigma^2_{Ni}$ are similar, only for the number of events in time
windows of length $\Delta t$. Under these conditions, it is also
trivial, that $\ev{f_i} = \ev{N_i}\ev{V_i}$.

All the above can be expected from simple principles. Two more
relationships are necessary and are often realized, they are basically the
same as \eqref{eq:s2V} and \eqref{eq:s2N}. The only strong assumption to
account for non-universal values of $\alpha$ is the following. Consider a
system, where elements with higher average activity do not only experience
more events, but those are also larger. Let us assume scaling between
the two quantities:
$$\ev{V_i} \propto \ev{N_i}^\beta,$$
which is equivalent to \eqref{eq:vvsn}. Then, $\alpha$ can be expressed
\cite{eisler.internal}, by combining all the formulas, as 
\begin{equation}
\alpha = \frac{1}{2}\left(1+\frac{\beta}{\beta+1}\right ).
\label{eq:internal}
\end{equation}

In this general context, the property $\beta \not = 0$ can be called
\emph{event size inhomogeneity} \footnote{In Refs. \cite{eisler.internal, eisler.nikkei}, $V_i(n)$ is called \emph{impact} and such property is called \emph{impact inhomogeneity}. However, here we wish to avoid confusion with another financial term, price impact \cite{gabaix.powerlaw, plerou.powerlaw, farmer.powerlaw, farmer.whatreally}.}. The intermediate values $0.5<\alpha<1$ interpolate between the square root type of $\ev{N}\propto \sigma_N^{1/2}$ and the linear $\ev{V}\propto \sigma_V$, while the conditions ensure that scaling is preserved (compare Figs. \ref{fig:alphaexampleNYSE} and \ref{fig:conditions}(a)-(b)).

These conditions are satisfied exactly in a random walker model on complex networks \cite{eisler.internal}.
Consequently, its behavior is well described by \eqref{eq:internal}.
However, such arguments can also be applied to stock market trading dynamics
when $\Delta t \ll 1$ min to ensure the validity of \eqref{eq:s2N}. By
substituting the observed values of $\beta$, one finds the estimates
$\alpha^*_\mathrm{NYSE} = 0.69 \pm 0.03$ and $\alpha^*_\mathrm{NASDAQ} = 0.59 \pm 0.02$. The
actual values are $\alpha_\mathrm{NYSE}(\Delta t \rightarrow 0) = 0.68 \pm 0.02$ and
$\alpha_\mathrm{NASDAQ}(\Delta t \rightarrow 0) = 0.67 \pm 0.02$. The
agreement for the NYSE data is good, for NASDAQ it is only
approximate. Moreover, Eq. \eqref{eq:vvsn} only fits the data for
large enough stocks, while Eq. \eqref{eq:alpha} gives an excellent fit
over the whole range available. Therefore, this explanation is only
partial, however, it indicates that $\alpha > 0.5$ is to be
expected. This is a crucial point, because markets are so far the only
examples of an $0.5 < \alpha < 1$ system.

\subsection{Time scale dependence of $\alpha$}
\label{sec:correl}
Section \ref{sec:stylized} revealed, that the exponent $\alpha$ of stock market
activity fluctuations shows a strong dependence on the time window $\Delta t$.
This was previously attributed to the effect of
external factors \cite{eisler.non-universality}. On the time scale of minutes, news, policy changes, etc. have
no time to diffuse in the system. Thus, temporal fluctuations are dominated by
internal dynamics, $\alpha < 1$. By increasing $\Delta t$ to days or weeks, the
importance of this external influence grows and $\alpha$ approaches $1$, which
is characteristic in the presence of a strong external driving. However, the
effect just described is a crossover, while observations show the persistence
of scaling, only the exponent $\alpha$ changes. This section offers an
alternative description that has no such shortcoming.

The key is to extend the analysis to $H(i)\not = 0.5$ systems. We start from
the relations \eqref{eq:hurst} and \eqref{eq:alpha},
where the role of the two variables $\ev{f_i}$ and $\Delta t$ is
analogous. When they hold simultaneously, from the equality of their left hand
sides, one can write the third proportionality
$$\Delta t^{H(i)} \propto \ev{f_i}^{\alpha(\Delta t)}.$$
After taking the logarithm of both sides, differentiation
$\partial^2/\partial(\log \Delta t)\partial(\log \ev{f_i})$ yields the
asymptotic equality
\begin{equation}
\gamma_{t} \sim \frac{d H(i)}{d (\log\ev{f_i})}\sim\frac{d \alpha(\Delta t)}
{d (\log\Delta t)}\sim \gamma_{f}.
\label{eq:gammagener}
\end{equation}
This means that both partial derivatives are constant and they have the same value,
which we will denote by $\gamma=\gamma_{t}=\gamma_{f}$.

\begin{figure*}[tb]
\centerline{\includegraphics[height=120pt]{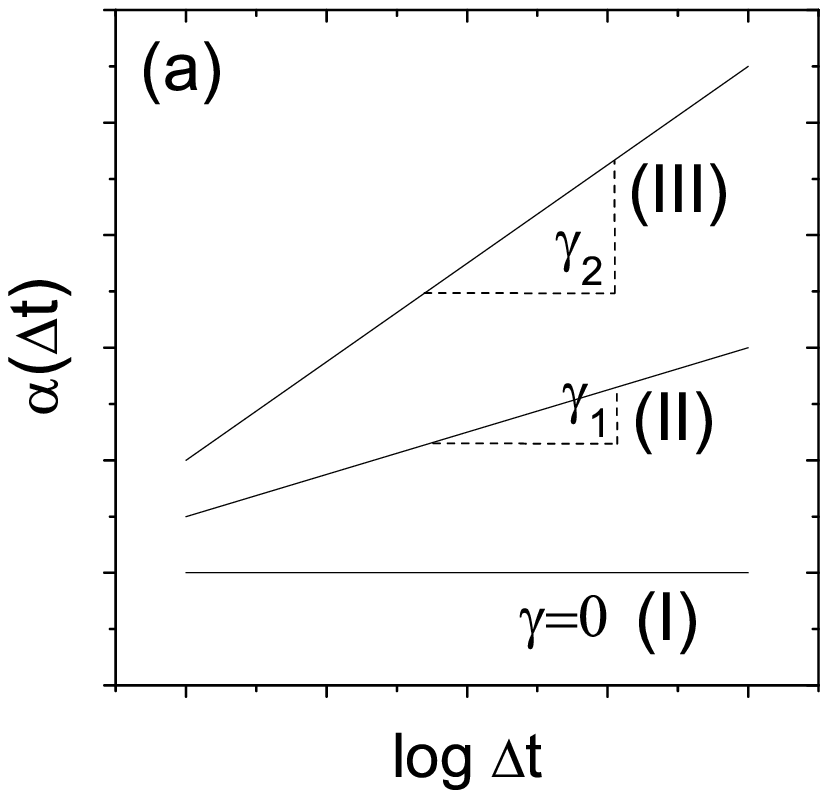}\hskip5mm
\includegraphics[height=120pt]{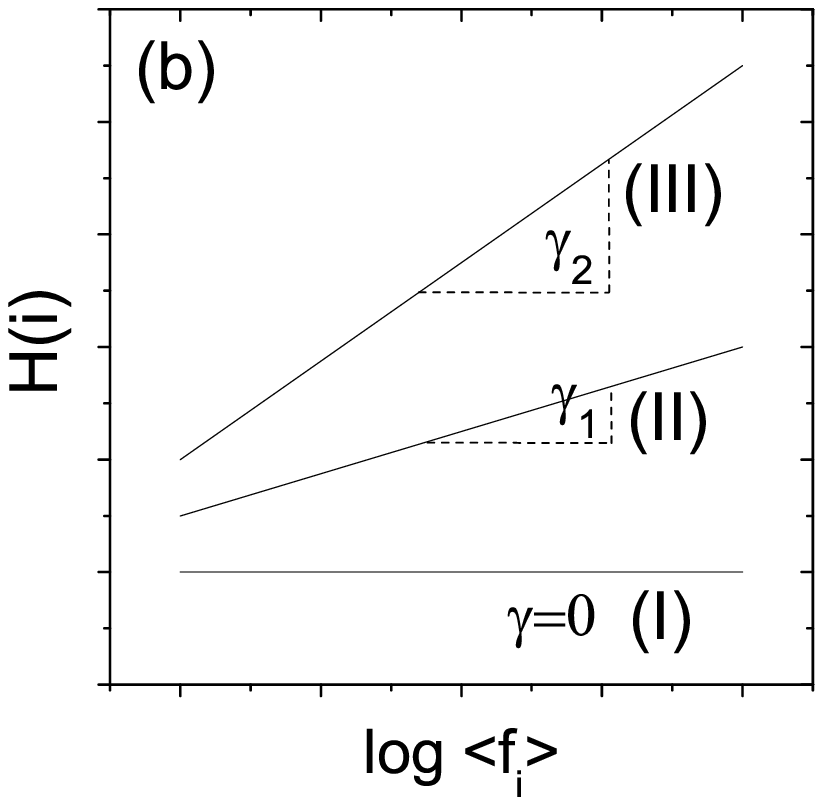}\hskip5mm
\includegraphics[height=120pt]{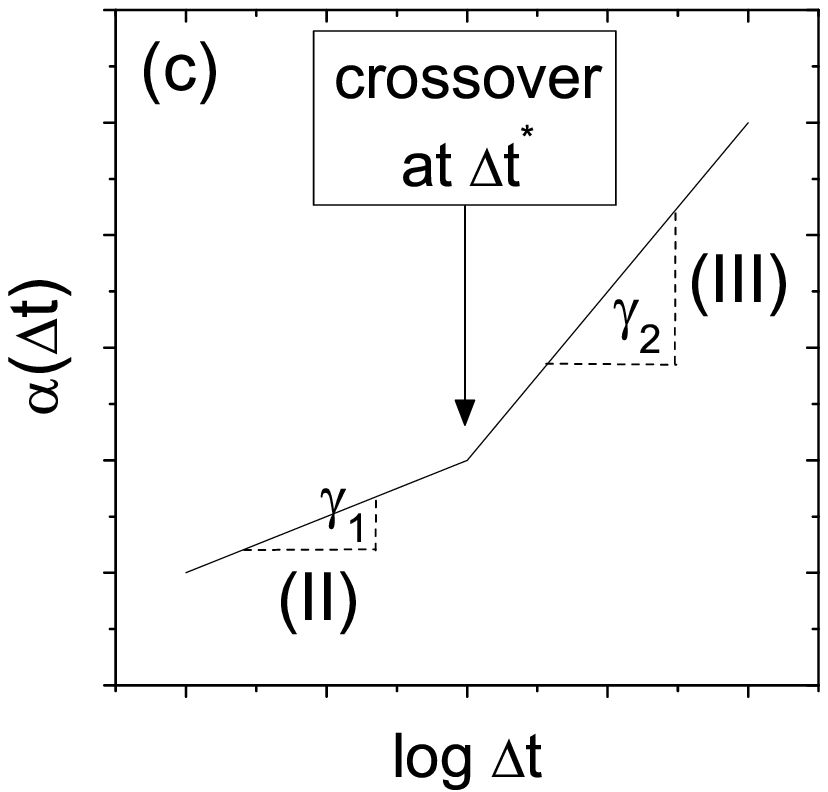}}
\caption{{\bf (a-b)} Possible scenarios where both
$\sigma_i(\Delta t)\propto \Delta t^{H(i)}$ and
$\sigma_i(\Delta t) \propto \ev{f_i}^{\alpha(\Delta t)}$ can be satisfied
simultaneously. (I) In systems, where $\gamma = 0$, $\alpha$ is independent of
window size and $H$ is independent of node. (II) When $\gamma = \gamma_1 > 0$,
$\alpha(\Delta t)$ and $H(i)$ depend logarithmically on $\Delta t$ and on
$\ev{f_i}$, respectively, with the common slope $\gamma_1$. (III) For a larger
value, $\gamma = \gamma_2 > \gamma_1$, the dependence is stronger.
{\bf (c)} Example of a crossover between different values of $\gamma$.
There, $\alpha$ still depends on $\Delta t$ in a logarithmic way, but the slope
is different in two regimes. In this case, for every node there are two Hurst
exponents, $H_1(i)$ and $H_2(i)$, that are valid asymptotically, for
$\Delta t \ll \Delta t^*$ and $\Delta t \gg \Delta t^*$, respectively. Then,
both of these must independently follow the logarithmic law shown in (b):
$H_1(i) = H_1^*+\gamma_1\log\ev{f_i}$ and $H_2(i) = H_2^*+\gamma_2\log\ev{f_i}$.}
\label{fig:sketch}
\end{figure*}

The possibilities how this can be realized are sketched in Figures
\ref{fig:sketch}(a)-(b):
\begin{enumerate}[(I)]
\item In systems, where $\gamma = 0$, the exponent
$\alpha(\Delta t) = \alpha^*$, it is independent of window size. At the same
time all nodes must exhibit the same degree of correlations, $H(i) = H$.
\item In the case, when $\gamma = \gamma_1 > 0$, $\alpha(\Delta t)$ actually
depends on $\Delta t$. This dependence must be logarithmic:
$\alpha(\Delta t) = \alpha^* + \gamma_1 \log \Delta t$. At the same time,
the Hurst exponent of the nodes depends on the mean flux in a similar way:
$H(i) = H^* + \gamma_1\log\ev{f_i}$. Moreover, the slope of the logarithmic
dependence is the same.
\item When the constant $\gamma$ is larger, for example $\gamma_2 > \gamma_1$
in Figures \ref{fig:sketch}(a)-(b), $\alpha$ changes faster with $\Delta t$,
while also $H(i)$ changes faster with $\ev{f_i}$.
\end{enumerate}
Finally, the combination of these options is also possible. Systems may display
a crossover between different values of $\gamma$ at a certain time scale
$\Delta t^*$, an example is given in Figures \ref{fig:sketch}(b)-(c).
There, $\alpha$ depends on $\Delta t$ in a logarithmic way, but the slope of
the trend is different in two regimes. In this case, there is no unique Hurst
exponent of $f_i(t)$. Instead, for every node there are two values, $H_1(i)$
and $H_2(i)$, that are valid asymptotically, for $\Delta t \ll \Delta t^*$ and
$\Delta t \gg \Delta t^*$, respectively. Then, both of these must independently
follow the logarithmic law: $H_1(i) = H_1^*+\gamma_1\log\ev{f_i}$ and
$H_2(i) = H_2^*+\gamma_2\log\ev{f_i}$.

Stock markets belong to this last group. For $\Delta t \leq 20$ min for NYSE and
$\Delta t \leq 2$ min for NASDAQ, $\alpha(\Delta t)\approx
\alpha^*$. Correspondingly, $H$ must be independent of $\ev{f}$, as it
was found in Section \ref{sec:stylized}. On the other hand, for $\Delta t
> 300$ min for NYSE and $\Delta t > 60$ min for NASDAQ, $\alpha
(\Delta t)$ is approximately logarithmic with the common coefficient
$\gamma = 0.06 \pm 0.01$. This, again, must equal the slope of $H(i)$
plotted versus $\log \ev{f_i}$. There is agreement between error bars
with the results of Section \ref{sec:stylized}.

The fact that the local derivative $\frac{d \alpha(\Delta t)}{d
(\log\Delta t)}$ also shows the degree of logarithmic trend in the
Hurst exponents, gives a visual method to detect the change in this
collective behavior of the market. Those regimes in $\Delta t$, where
$\alpha(\Delta t)$ is constant, correspond to time scales where all
stocks have the same level (Hurst exponent) of activity
correlations. Where $\alpha(\Delta t)$ is logarithmically changing,
the slope $\gamma$ gives the degree of inhomogeneity in $H(i)$. Finally,
the function is curved near crossovers, where the degree of the mean flux dependence
in correlation strengths is changing.

In order to underline, that the $\alpha(\Delta t)$ dependence comes
from temporal correlations, we carried out the same measurement,
but with all time series shuffled randomly. It is trivial, that if
$\Delta t$ equals the $\delta = 1$ sec resolution of the dataset,
shuffling does not affect the estimates of $\sigma_i(\Delta
t=\delta)$, it merely rearranges the terms used in averaging \footnote{In fact, the DFA procedure can only be
applied for $\Delta t\geq 4\delta$, but the effect of this difference
is negligible.}. Hence, the fitted slope cannot change either, $\alpha_\mathrm{shuff}(\delta)=\alpha(\delta)$.
On the other hand, shuffling
gives uncorrelated time series, $H_\mathrm{shuff}(i)\equiv0.5$ (see Section
\ref{sec:stylized}). Correspondingly, $\gamma_\mathrm{shuff} = \frac{d
H_\mathrm{shuff}}{d \log \ev{f}}= 0$. Hence, according to
\eqref{eq:gammagener}, $\alpha_\mathrm{shuff}(\Delta t) = \alpha^*$,
regardless of window size. The measurement results -- in excellent
agreement with the above reasoning -- are shown by empty circles in
Figs. \ref{fig:alpha}(a) and (b).

 Finally, we must emphasize that the value of $\gamma$ is not \emph{a priori} known for real systems. Consequently, $\alpha$ does not reflect the type of internal dynamics in the straightforward fashion suggested by Ref. \cite{barabasi.fluct}. Instead, a careful analysis, including the dependence on $\Delta t$, must be undertaken, in
order to interpret the results correctly. The only exception is when we can assume homogeneous correlations,
i.e., $\gamma = 0$ and so $H(i)=H$.

\section{Conclusions}

In the above, we generalized a fluctuation scaling relation to the
case when temporal correlations are present in the individual time series.
In such analysis, one measures the time-scale dependent scaling exponent
$\alpha(\Delta t)$. In addition to previous studies,
we found that even in the presence of strong temporal correlations, $\alpha$
still remains very characteristic to the internal dynamics. Indeed, its
time scale dependence reveals additional information. For the persistence
of fluctuation scaling at all time scales,
it is inevitable that the strength of correlations in \emph{all} the individual
time series be connected by a logarithmic law. Such a relationship is a
peculiar feature of collective dynamics, which is not explained by the
number or the size distribution of the events that occur at the nodes.

The framework was applied to reveal the connections between stylized facts
of stock market trading activity. Empirical data for both of the markets NYSE
and NASDAQ show qualitatively similar behavior. The values of $\alpha(\Delta t)$ 
can be understood based on the role of company size. For short times when
there are no correlations between the trades of an individual company,
the non-trivial value of $\alpha$ comes from the highly inhomogeneous
trade sizes of the different companies. For increasing time windows,
we observe a logarithmic law in correlation strengths and that this
leads to a window size dependence of $\alpha$. As the growing size of
individual trades with increasing company size can also be considered as a
cumulation of smaller transactions, our results underline the importance
of temporal correlations and size dependence in explaining scaling phenomena
on the stock market.

\section*{Acknowledgments}

The authors are indebted to Gy\"orgy Andor for his support with financial data.
They also thank an anonymous referee for suggestions, including ways to improve the
clarity of the manuscript. JK is member of the Center for Applied Mathematics and Computational
Physics, BME. Support by OTKA T049238 is acknowledged.

\bibliographystyle{unsrt}
\bibliography{unifieda}

\end{document}